\documentclass[aps,prl,showpacs,preprint,endfloats*]{revtex4}

\usepackage[dvips]{graphicx}
\usepackage{amsmath}

\newcommand{\mterm}[4]{\mbox{$\mathrm{#1}^{#2}\Sigma^{#3}_{#4}$}}
\newcommand{\um}{\mbox{$\mu\mathrm{m}$}}

\newcommand{\us}{\mbox{$\mu\mathrm{s}$}}

\begin{document}

\title{Photoassociation of sodium in a Bose-Einstein
condensate} 

\author{C.~M$\mathrm{^c}$Kenzie}
\author{J.~Hecker~Denschlag}
\altaffiliation{
Institut f\"{u}r Experimentalphysik,
Universit\"{a}t Innsbruck,
Technikerstrasse 25,
A-6020 Innsbruck, Austria.
}
\author{H.~H\"affner} 
\altaffiliation{
Institut f\"{u}r Experimentalphysik,
Universit\"{a}t Innsbruck,
Technikerstrasse 25,
A-6020 Innsbruck, Austria.
}
\author{A.~Browaeys} 
\author{Lu\'\i s~E.~E.~de~Araujo} 
\altaffiliation{
The Institute of Optics, University of Rochester, Rochester, NY
14627, USA.
}
\author{F.~K.~Fatemi}
\altaffiliation{
Naval Research Lab, Washington, DC 20375, USA.
}
\author{K.~M.~Jones}
\altaffiliation{
Department of Physics, Williams College, Williamstown, MA 01267, USA.
}
\author{J.~E.~Simsarian}
\altaffiliation{
Bell Laboratories,
Lucent Technologies,
Holmdel, NJ 07733, USA.
}
\author{D.~Cho}
\altaffiliation{
Dept. of Physics, Korea University, 5-1 Ka Anam-dong, Sungbuk-ku, Seoul 136-701, Korea.
}
\author{A.~Simoni}
\altaffiliation{
INFM and LENS, Universit\`{a} di Firenze, Largo E. Fermi 2, I-50125
Firenze, Italy.
}
\author{E.~Tiesinga}
\author{P.~S.~Julienne}
\author{K.~Helmerson}
\author{P.~D.~Lett}
\author{S.~L.~Rolston}
\author{W.~D.~Phillips} 

\affiliation{National Institute of Standards and
Technology, Gaithersburg, MD 20899, USA.} 
\date{Preprint \today}

\begin{abstract}
We report on the formation of ultra-cold Na$_2$ molecules using
single-photon photoassociation of a Bose-Einstein condensate. The
photoassociation rate, linewidth and light shift of the $J=1$, $v=135$
vibrational level of the \mterm{A}{1}{+}{u}\ molecular bound state
have been measured. We find that the photoassociation rate constant
increases linearly with intensity, even where it is predicted that
many-body effects might limit the rate. Our observations are
everywhere in good agreement with a two-body theory having no free
parameters.
\end{abstract}


\maketitle

Bose-Einstein condensates (BECs) of atomic gases
\cite{Anderson1995,Davis1995} are versatile systems for the
experimental study of quantum behavior. Of particular interest are the
suggestions for the coherent coupling of a BEC of atoms with a BEC of
molecules \cite{Heinzen2001,Kostrun2000,Timmermans1999,Julienne1998}
and the possibility of creating entangled pairs of atoms in the BEC
via coupling with molecular levels \cite{Helmerson2001}. Two-photon
photoassociation processes using stimulated Raman transitions have
formed ground state molecules from ground state atoms in a BEC
\cite{Wynar2000,Gerton2000}\, but at very low rates. Here we explore
the fundamental upper limits of molecule formation by making them at
high rates using the elementary process of single-photon
photoassociation.

In single-photon photoassociation two atoms collide in the presence of
a light field and form an excited state molecule. Photoassociative
spectroscopy is used extensively to study collisions between laser
cooled atoms \cite{Weiner1999}. Photoassociation in a BEC presents
quite a different regime: the collision energies are orders of
magnitude lower than in a laser cooled sample (the de Broglie
wavelength is as big as the sample) and the densities are higher.
This puts us in a regime where many-body effects may be important.

We concentrate on a particular photoassociation transition and measure
the photoassociation spectra for various intensities and durations of
the light pulse. From these we determine the photoassociation rate,
lineshape and the shift of the resonance. Finally we examine various
limits on the photoassociation rate. 

\begin{figure}
\includegraphics[width=7cm]{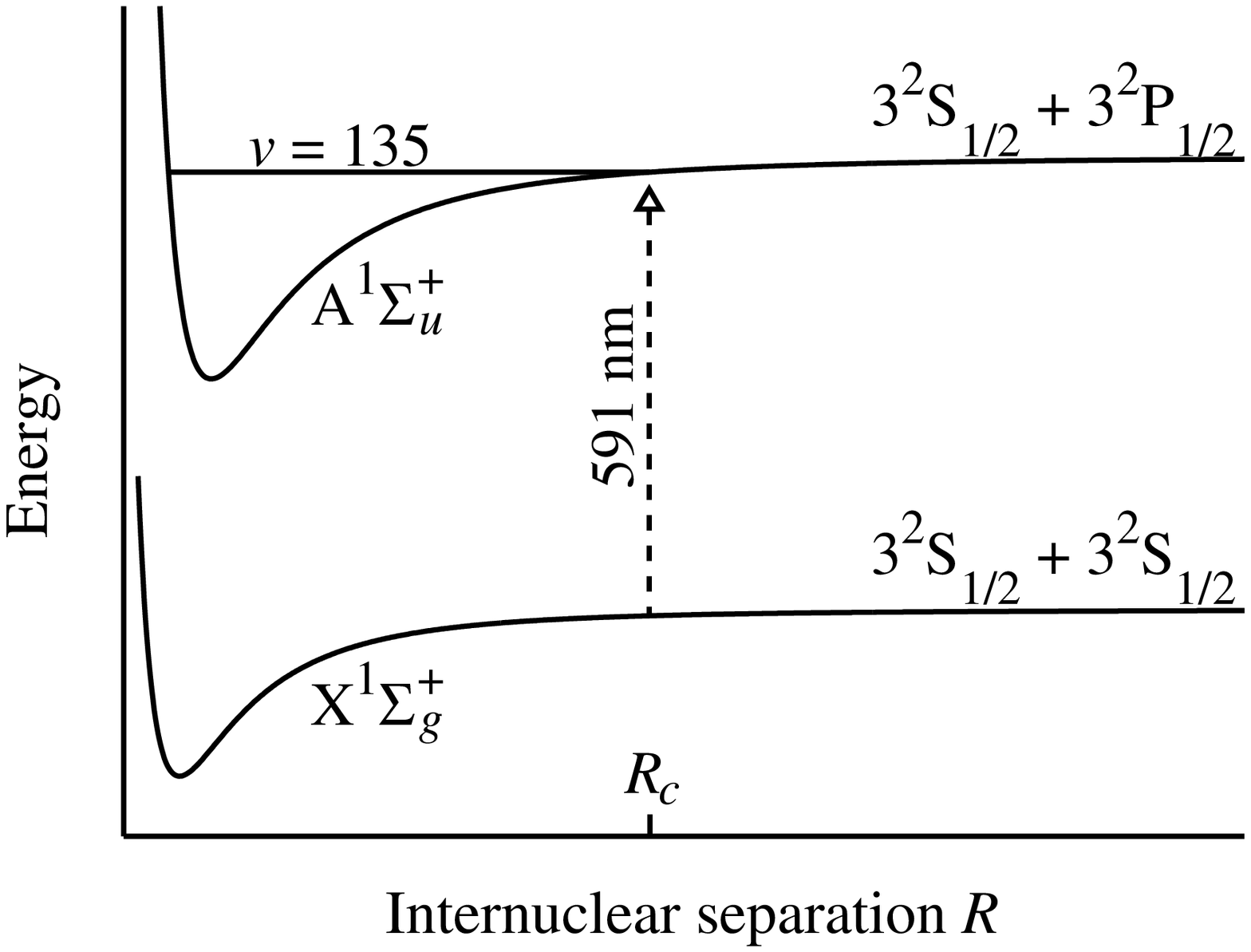}%
\caption{The two-atom potentials for the ground state and the excited
state used for photoassociation. The atoms are initially unbound and
on the ground state asymptote and are excited into the $J=1$, $v=135$
level. From there they decay and are lost from the condensate. $R_C =
2.0$ nm is the Condon radius, the internuclear separation where the
energy of a resonant photon matches the difference between the potentials.}
\label{fig:levels}
\end{figure}

Figure \ref{fig:levels} shows the photoassociation process.  The
molecular level chosen for study is the $J=1$, $v=135$,
rotational-vibrational level of the \mterm{A}{1}{+}{u}\ Na$_2$
molecular state, excited from free atoms by a laser frequency of
16913.37(2) cm$^{-1}$ \footnote{Unless otherwise stated, all
uncertainties reported here are one standard deviation combined
systematic and statistical uncertainties.}. We chose this level
because its detuning from the D$_1$ resonance is far enough (43
cm$^{-1}$) for atomic absorption to be negligible and because earlier
experiments using laser cooled atoms indicated a high photoassociation
rate. The lifetime of our excited molecules is about 8.6 ns. Decay of
excited molecules into hot atoms or ground state molecules constitutes
loss from the condensate. This loss is how we detect photoassociation.

We prepare an almost pure condensate of $N\approx 4\times10^6$ sodium
atoms in the $|F=1, m_F=-1\rangle$ ground state with a peak density of
$n_0\approx 4\times 10^{14}$ cm$^{-3}$. The condensate is held in an
anisotropic magnetic TOP \cite{Kozuma1999} trap with oscillation
frequencies of $\omega_x/\sqrt{2} = \omega_y = \sqrt{2}\omega_z$
= $2\pi\times 198$~Hz and corresponding Thomas-Fermi radii of
$\sqrt{2}r_x = r_y = r_z/\sqrt{2} = 15$ \um.

To induce photoassociation we illuminate the condensate with a
Gaussian laser beam focused to 120 \um\ FWHM at the condensate. The
peak intensity is varied from 50 to 1200 W$\cdot$cm$^{-2}$. The
polarization is linear and parallel with the rotation ($z$)
axis of the TOP trap bias field. The light is applied as a square
pulse for between 10 and 400 \us\ with rise and fall times of less
than 0.5 \us.

The condensate number is measured using phase contrast imaging
\cite{Andrews1996}, taking two images before and two images after the
photoassociation pulse to determine loss. The imaging occurs at 40 ms
intervals using a 100 $\mu$s probe pulse from a laser tuned 1.78 GHz to
the red of the 3S$_{1/2}, F=1 \rightarrow\:3$P$_{3/2}, F=2$
transition. The imaging rate is limited by the readout time of the CCD
camera. The photoassociation pulse begins halfway between the second
and third images. We use multiple imaging pulses to improve statistics
and to partially correct for small losses other than those due to the
photoassociation pulse. These losses are typically 4\% between
images. Once the number of atoms is extracted from the images
\footnote{Because we are in a regime where the phase shift of the
light passing through the BEC is large (up to several radians) the
extracted density is a multi-valued function of the phase contrast
signal. We assume a Thomas-Fermi distribution for the atoms to resolve
this ambiguity.} we calculate $f$, the fraction of atoms remaining
after the photoassociation pulse.

\begin{figure}
\includegraphics[width=7cm]{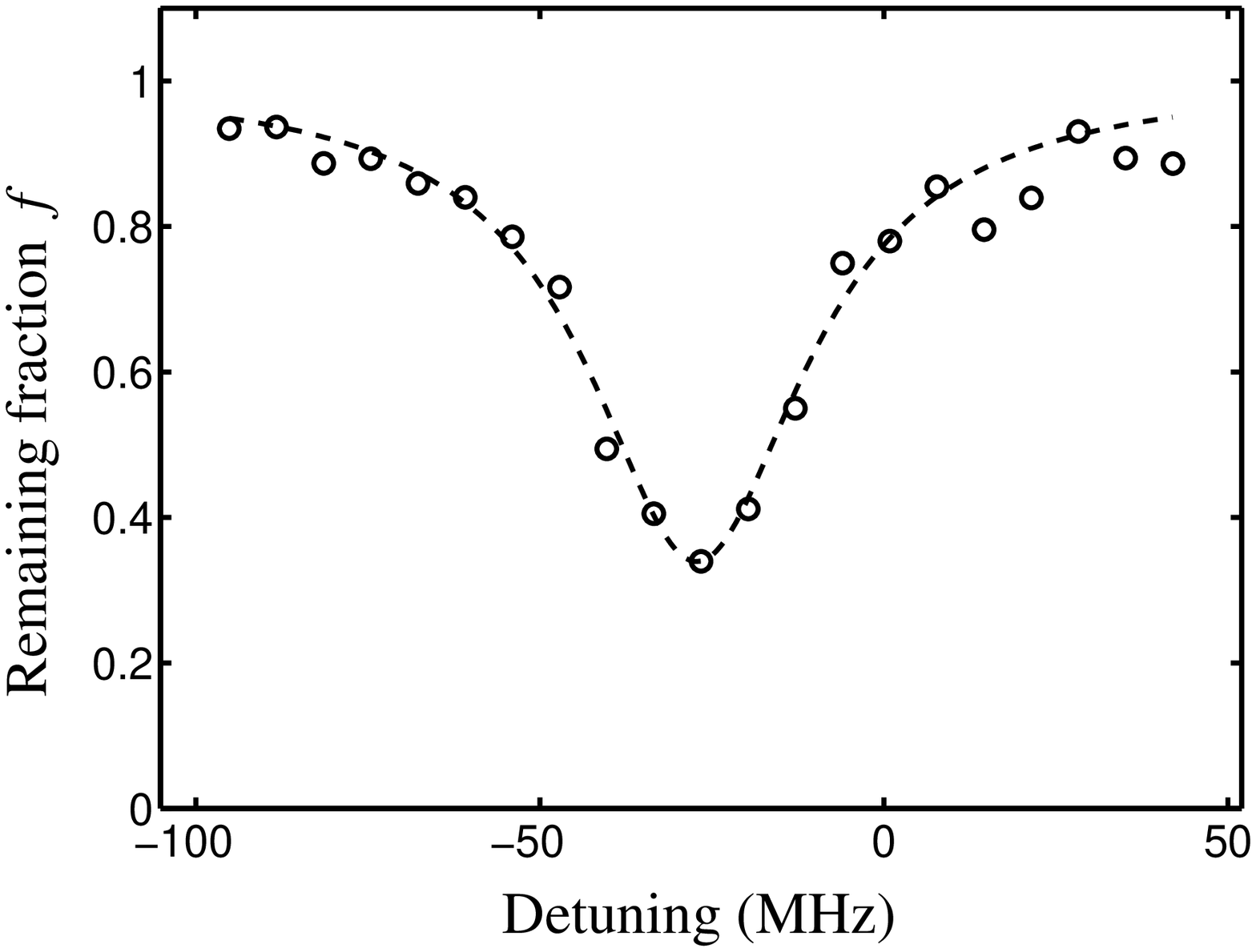}%
\caption{A typical photoassociation loss spectrum. A
140~W$\cdot$cm$^{-2}$ pulse was applied for 100 $\mu$s. The dotted
line is a fit to equation~(\ref{equ:global}). The uncertainty in the
frequency for each point is 5MHz.}
\label{fig:spectrum}
\end{figure}

Figure~\ref{fig:spectrum} shows a typical photoassociation
spectrum. Each point represents a freshly prepared condensate. We use a
Fabry-Perot etalon and a reference laser locked to an atomic Na line
to measure differences in the photoassociation frequency with a
precision of 5 MHz. The laser linewidth is $<3$ MHz. All detunings
quoted are relative to the center of the photoassociation line in the
low intensity limit. For small trap loss we expect the line to be
Lorentzian (in contrast to photoassociation lines in an uncondensed
thermal sample where the kinetic energy distribution distorts the line
shape \cite{Jones2000}). For significant trap loss, as in
figure~\ref{fig:spectrum}, one must account for the change of the
density profile during the photoassociation pulse.

\begin{figure}
\includegraphics[width=7cm]{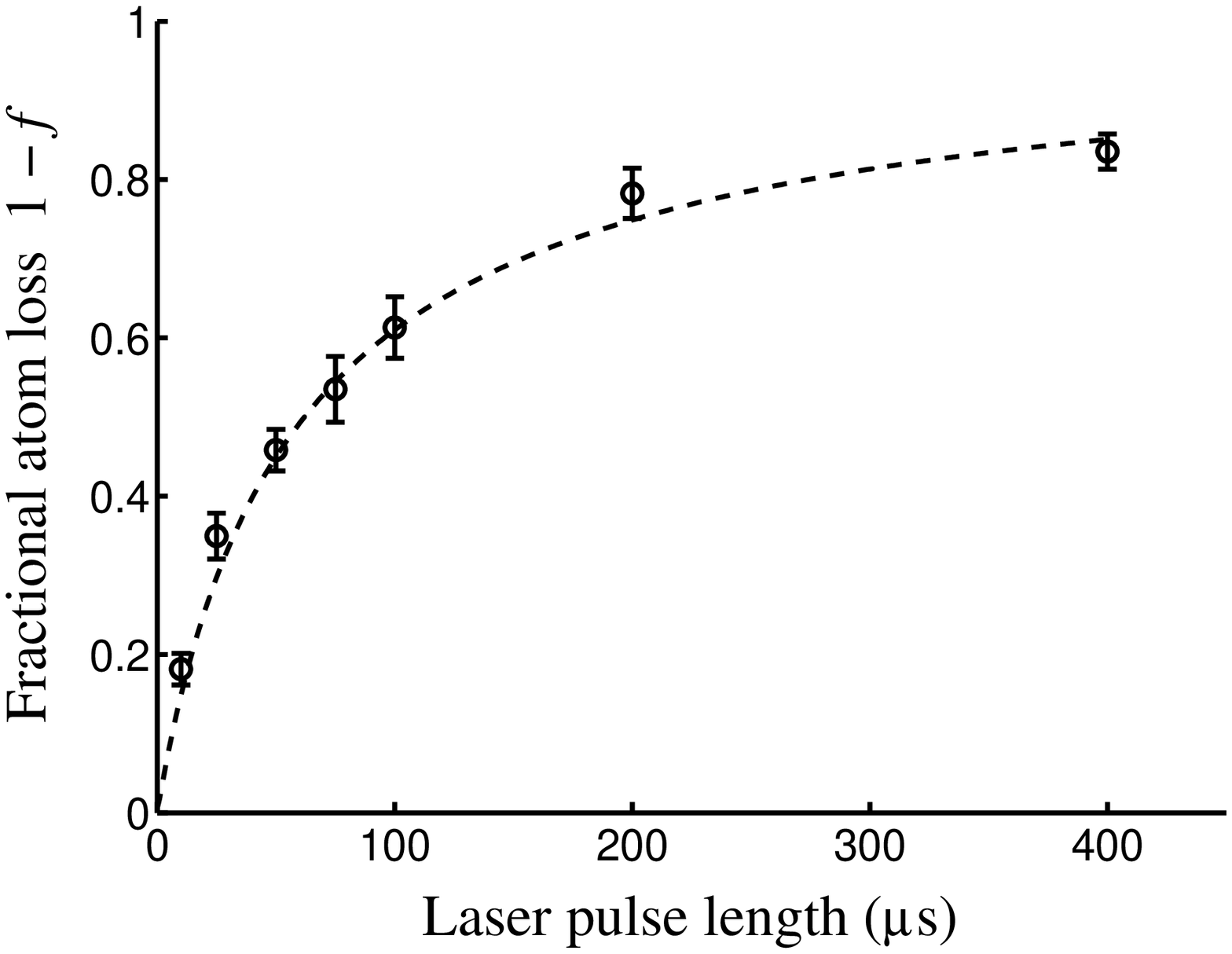}
\caption{The dependence of the maximum condensate loss on the photoassociation
pulse length for $I = 140$ W$\cdot$cm$^{-2}$; the curve is a fit of equation
(\ref{equ:global}).}
\label{fig:timedep}
\end{figure}

A two-body collisional loss process changes the local atomic density
as: $\dot{n}(t,{\mathbf r}) = -K(I,\omega)n^2(t,{\mathbf r})$, where
$K(I,\omega)$ is the intensity and frequency dependent photoassociation
rate constant.  Because the characteristic time for the motion of the
atoms, the trap oscillation period, is long compared to the
photoassociation pulse, we assume that the local density only changes
due to photoassociation. We then obtain:
\begin{equation}
\label{equ:local}
n(t,{\mathbf r}) = \frac{n(0,{\mathbf r})}{1+K(I,\omega)n(0,{\mathbf r})t}.
\end{equation}
The density distribution flattens with time. Spatially integrating
equation~(\ref{equ:local}), assuming an initial, parabolic
(Thomas-Fermi), density distribution and a uniform intensity $I$, leads to
an expression for the fraction of atoms remaining in the condensate
\begin{multline}
\label{equ:global}
f(\eta) = \frac{15}{2}\eta^{-5/2}\left\{ \eta^{1/2} + \frac{1}{3}\eta^{3/2}\right.\\
\left.-\;(1+\eta)^{1/2}\tanh^{-1}(\sqrt{\eta/(1+ \eta)})
\right\},
\end{multline}
where $\eta = K(I,\omega)n(0,{\mathbf 0})t = K_m(I)n(0,{\mathbf
0})t/(1 + (2(\omega - \omega_0(I))/\gamma(I))^2)$.  We use a three
parameter fit of equation~(\ref{equ:global}) to the spectra to extract
the on-resonance rate constant $K_m(I)$, effective line width
$\gamma(I)$ and central frequency $\omega_0(I)$ (for example, the
dotted line in figure~\ref{fig:spectrum}). The fit is good. To
further verify equation~(\ref{equ:global}) we plot the measured
$1-f$ as a function of pulse length for $I = 140$
W$\cdot$cm$^{-2}$ and $\omega = \omega_0$, along with a one parameter
($K_m$) fit to the data (figure~\ref{fig:timedep}). The error bars are
the fitting uncertainty.

By fitting spectra obtained at various intensities we measure
$K_m(I)$, $\gamma(I)$, and $\omega_0(I)$. We calculate the
unbroadened molecular linewidth of the chosen state to be $\gamma_0/2\pi =
18.5$ MHz (nearly twice the atomic linewidth). This is in good
agreement with the measured linewidth of 19.5(25)
MHz in the low intensity limit where it is independent of
intensity. At higher intensities we observe broadening with a maximum
linewidth of 60~MHz at 1~kW$\cdot$cm$^{-2}$. Homogeneous power
broadening is calculated to be three orders of magnitude too low to
explain this width. It is, however, partially explained by
differential light shifts of the unresolved molecular hyperfine
states. These states are calculated to be split by less than 1 MHz at
low intensities and about 30 MHz at our maximum intensity. Another
possible contribution is the inhomogeneity of the photoassociation
beam intensity combined with the large light shift (discussed
below). Variations due to either local spatial inhomogeneity
(e.g. interference fringes) or displacement of the cloud from the
center of the Gaussian beam could account for the extra
width. Assuming these inhomogeneous broadening mechanisms do not
change the area of the line (verified by a simulation), we take the
on-resonance photoassociation rate constant to be $K_0(I) =
K_m(I)\gamma(I)/\gamma(I\rightarrow 0)$.

\begin{figure}
\includegraphics[width=7cm]{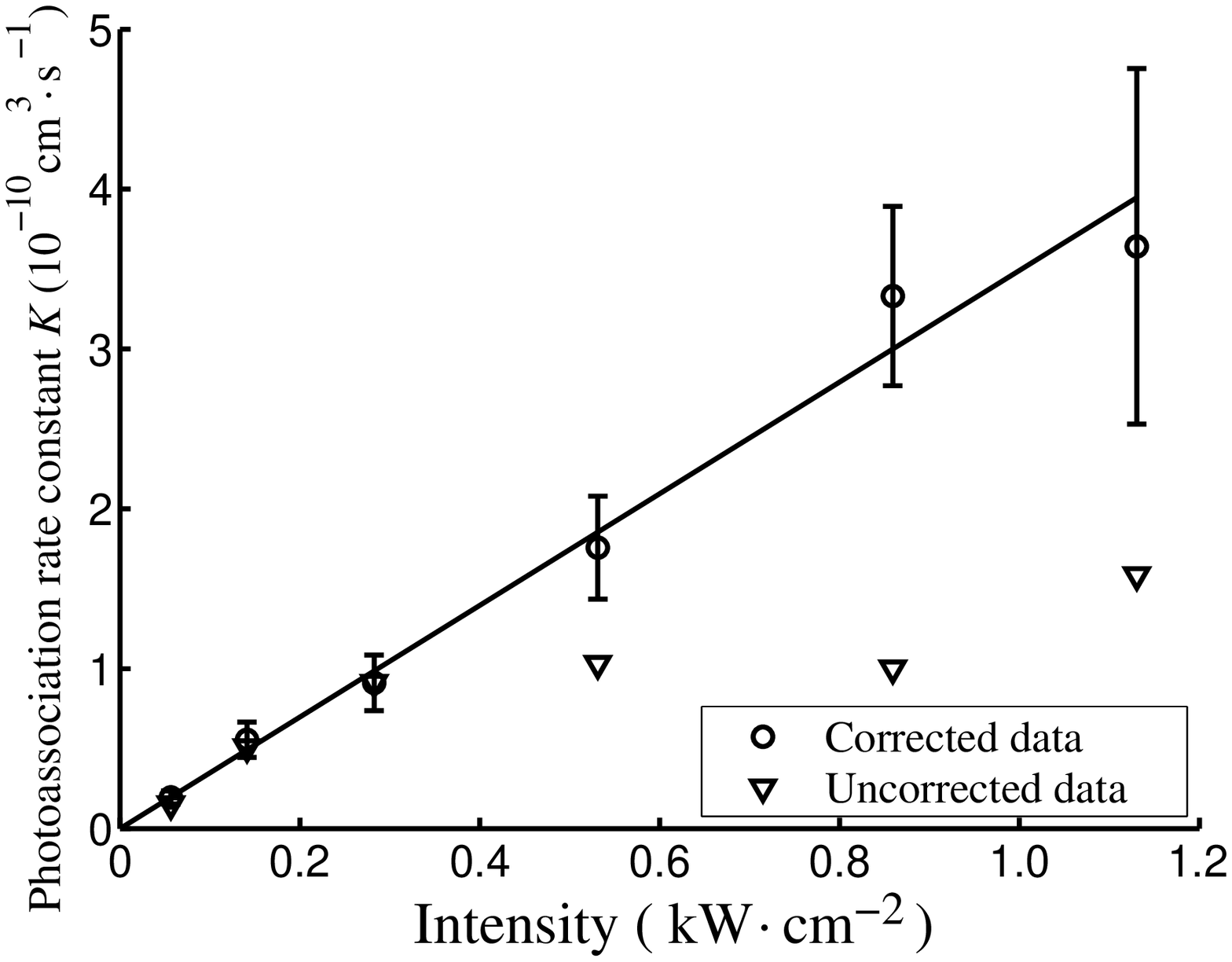}
\caption{Photoassociation rate constant as a function of
intensity. The corrected data has been adjusted to account for the
inhomogeneous broadening.}
\label{fig:intdep}
\end{figure}

Figure \ref{fig:intdep} shows corrected and uncorrected $K$ as a
function of intensity (for various pulse lengths). The error bars are
the fitting uncertainties. Once we correct for the inhomogeneous
broadening we get a linear dependence on intensity with a slope of
$dK_0/dI = 3.5(2)(10)\times10^{-10}\;
(\mathrm{cm}^3\cdot\mathrm{s}^{-1}) /
(\mathrm{kW}\cdot\mathrm{cm}^{-2})$. The uncertainties are due,
respectively, to fitting and to the combined uncertainties in the
measurement of the intensity and density. For intensities above
1.2~kW$\cdot$cm$^{-2}$, which we could only achieve by more tightly
focusing the photoassociation laser, atomic dipole forces
significantly perturb the condensate, thwarting meaningful
measurements. A coupled-channels, two-body scattering calculation with
no adjustable parameters \cite{Samuelis2001} yields a photoassociation
rate constant of $dK_0/dI = 4.1\times10^{-10}\;
(\mathrm{cm}^3\cdot\mathrm{s}^{-1}) /
(\mathrm{kW}\cdot\mathrm{cm}^{-2})$ for our range of intensities. This
includes a factor of 2 decrease relative to $K$ for a non-condensed
gas and agrees well with our experimental result.

\begin{figure}
\includegraphics[width=7cm]{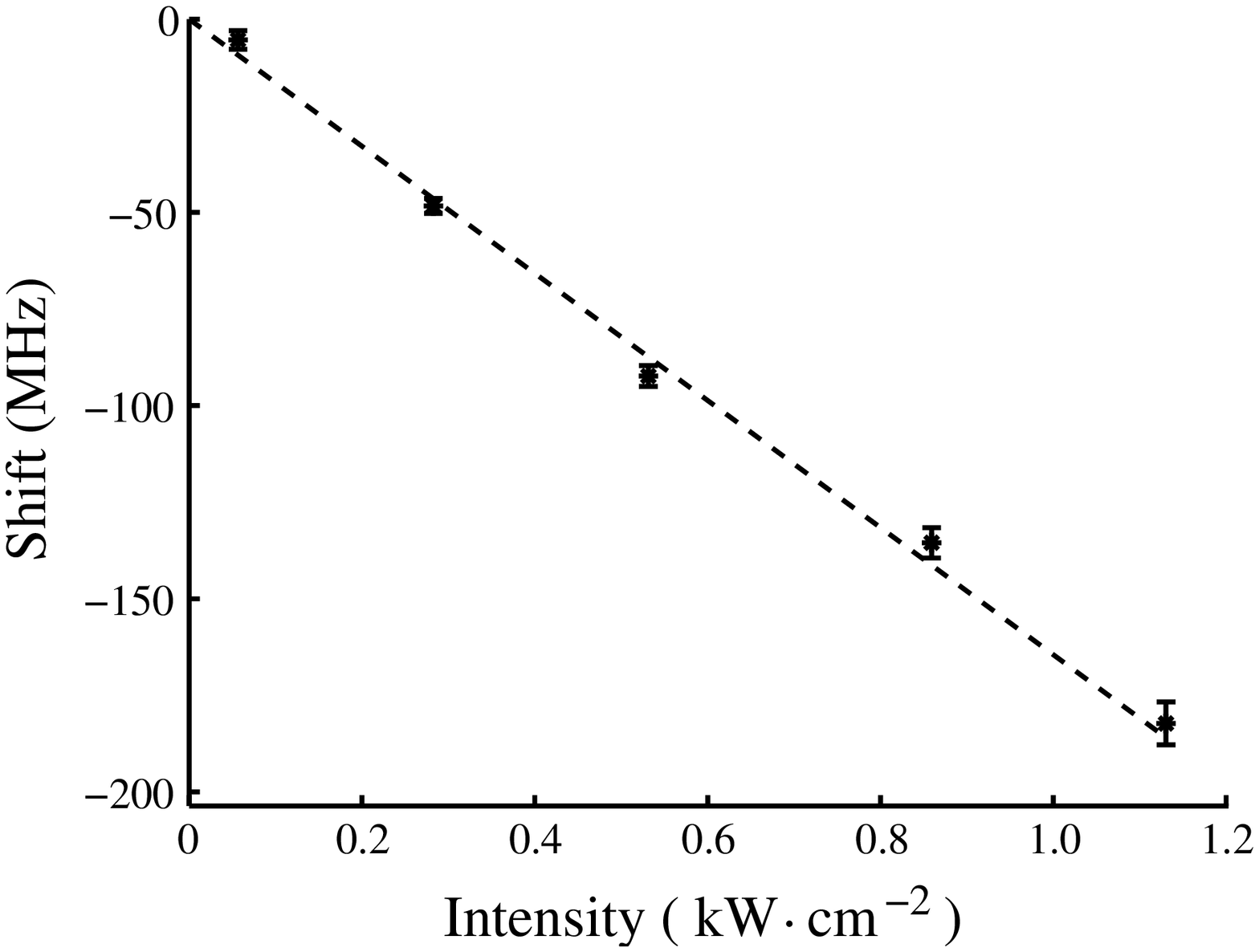}
\caption{The light shift of the resonance as a function of laser
intensity.}
\label{fig:lightshift}
\end{figure}

We study the photoassociation light shift, previously observed in a
non-condensed gas \cite{Jones1997}, in a set of experiments where the
total fluence (intensity $\times$ pulse length) of the pulse was kept
constant, to maintain the depth of the photoassociation dip in an
easily observable regime. The results are shown in figure
\ref{fig:lightshift}. The measured light shift is
$-164(35)\:\mathrm{MHz}\left
/\frac{\mathrm{kW}}{\mathrm{cm}^2}\right.$,
 which leads at high intensity to a shift large compared to the
linewidth. The principal contribution to the uncertainty is the
intensity calibration. During preparation of this work we
became aware of similar results in lithium \cite{Gerton2001}.

While the strength of the photoassociation resonance is dominated by
s-wave scattering, the dominant contribution to the light shift is due
to a d-wave shape resonance. A theoretical calculation of the light
shifts using equation (3.7) of reference \cite{Bohn1999},
including the effect of the d-wave shape resonance embedded in the
continuum, gives an average value of
$-130$~$\mathrm{MHz}\left/\frac{\mathrm{kW}}{\mathrm{cm}^2}\right.$ with
a spread of
$\pm$13~$\mathrm{MHz}\left/\frac{\mathrm{kW}}{\mathrm{cm}^2}\right.$
due to the hyperfine structure.

We now consider the upper limit to the photoassociation rate constant
K [this implies a lower limit on the photoassociation time $\tau =
(Kn)^{-1}$]. If one uses a semi-classical theory that is commonly
applied to collisions of laser-cooled atoms \cite{Gallagher1989} then
$K_0=\sigma v=\pi R_C^2Pv$, where $R_C$ is the Condon radius (see
figure \ref{fig:levels}) and $P$ is the probability of a
photoassociative transition with a maximum value of 1. If we take the
relative velocity $v$ of the atoms to be $h/(2mr_y) = 0.6$
mm$\cdot$s$^{-1}$, where $m$ is the atomic mass, then the maximum
photoassociation rate constant is four orders of magnitude lower than
our highest measured value. This reveals the inadequacy of a
semi-classical approach, which fails to take into account threshold
laws \cite{Weiner1999}.

Quantum theories for the photoassociation rate constant can be
compared by expressing $K$ as $K=(h/m)L$, where $L$ is a
characteristic length. Two-body s-wave scattering theory for a BEC
gives $L_s = |S(k)|^2/k$ where $\hbar k$ is the relative collision
momentum and $S(k)$ is the $S$-matrix element for atom loss via
photoassociation. References \cite{Bohn1996,Bohn1999} show that, on
resonance,
\begin{equation}
    L_s(I) = \frac{4\gamma_0 \Gamma(k,I)/k}{(\gamma_0 +
    \Gamma(k,I))^2}
    \label{Bpeak}
\end{equation}
where $\hbar \Gamma(k,I) = 2 \pi |\langle e| \hbar \Omega |
k\rangle|^2$ is the Fermi-golden-rule stimulated-decay width of the
excited molecular state $|e\rangle$ due to the optical coupling $\hbar
\Omega \propto \sqrt{I}$ with the colliding atoms. Since $\Gamma
\propto k$ as $k\rightarrow 0$, and $\Gamma/\gamma_0 < 0.001$ in our
range of power and collision energy, $L_s$ is independent of $k$. We
calculate $\gamma_0/2\pi = 18.5$ MHz.  $L_s$ is linear in $I$ for our
experimental conditions and $dL_s/dI$ is calculated to be 24
nm$/\frac{\mathrm {kW}}{\mathrm{cm}^2}$.  This gives the above-quoted
rate constant in good agreement with the experiment.  In our power
range, $L_s$ can be significantly larger than the Condon point for the
transition, 2.0 nm. Note that equation~(\ref{Bpeak}) shows that
$L_s(I)$ will saturate with increasing $I$ and decrease for
sufficiently large $I$.

The upper limit to the two-body quantum $K$ is the unitarity limit
where $|S|^2 = 1$ so $L_u = 1/k = \lambda/2\pi$, where $\lambda$ is
the de Broglie wavelength. Since $\lambda$ is on the
order of the BEC size $L_s/L_u \ll 1$ so that our
experiment is well below the unitarity constraint. 

Recent many-body theoretical
work~\cite{Kostrun2000,Javanainen2001,Javanainen2001b}
has suggested an upper limit to $K$ in a BEC of $K \sim \frac{h}{m}
L_J$, where $L_J = \frac{n^{-1/3}}{2\pi}$ and $n^{-1/3}$ is the mean
distance between particles. One might question if two-body
scattering methods are applicable at densities where $L_s$ becomes
larger than $L_J$. At our maximum density $L_J = 22$ nm, so
$L_s/L_J \approx 1$ at our highest intensities. This is the regime
where one might expect two-body theory to start to
fail. Nevertheless the linearity of $K(I)$ at our highest intensities
(figure~\ref{fig:intdep}) suggests that two-body theory continues to
be valid.

Larger values of $L_s/L_J$ might be accessible by a modification to
our experimental design. We can use the atomic dipole force (which
currently limits our ability to use high intensities) to our advantage
by trapping the atoms with the photoassociation laser. Without
changing the atomic dipole forces, the laser can be suddenly brought
from far off molecular resonance to on molecular resonance to induce
photoassociation. Difficulties due to the molecular light shift might
be reduced by finding a transition with a smaller light shift.

In conclusion, we have measured the single-photon photoassociation in a
BEC, in good agreement with two-body theory. This agreement represents
a confirmation of the factor-of-two reduction for a two-body
inelastic process in a BEC. The characteristic time for
photoassociation is as short as 5 $\mu$s, much shorter than the 100
$\mu$s to traverse the mean distance between atoms, another
demonstration of the extreme quantum nature of the collisions. Our
largest rate is still much smaller than the unitarity limit, but is about
equal to a limit suggested on the basis of many-body effects;
we do not see evidence that the rate saturates at this limit.

\begin{acknowledgments}
We acknowledge funding support from the US Office of Naval Research
and NASA. J.H.D. and H.H. acknowledge funding from the Alexander von
Humboldt foundation. A.B. was partially supported by DGA (France).
\end{acknowledgments}

\bibliography{molecules}

\begin{thebibliography}{21}
\expandafter\ifx\csname natexlab\endcsname\relax\def\natexlab#1{#1}\fi
\expandafter\ifx\csname bibnamefont\endcsname\relax
  \def\bibnamefont#1{#1}\fi
\expandafter\ifx\csname bibfnamefont\endcsname\relax
  \def\bibfnamefont#1{#1}\fi
\expandafter\ifx\csname citenamefont\endcsname\relax
  \def\citenamefont#1{#1}\fi
\expandafter\ifx\csname url\endcsname\relax
  \def\url#1{\texttt{#1}}\fi
\expandafter\ifx\csname urlprefix\endcsname\relax\def\urlprefix{URL }\fi
\providecommand{\bibinfo}[2]{#2}
\providecommand{\eprint}[2][]{\url{#2}}

\bibitem[{\citenamefont{Anderson et~al.}(1995)\citenamefont{Anderson, Ensher,
  Matthews, Wieman, and Cornell}}]{Anderson1995}
\bibinfo{author}{\bibfnamefont{M.~H.} \bibnamefont{Anderson}},
  \bibinfo{author}{\bibfnamefont{J.~R.} \bibnamefont{Ensher}},
  \bibinfo{author}{\bibfnamefont{M.~R.} \bibnamefont{Matthews}},
  \bibinfo{author}{\bibfnamefont{C.~E.} \bibnamefont{Wieman}},
  \bibnamefont{and} \bibinfo{author}{\bibfnamefont{E.~A.}
  \bibnamefont{Cornell}}, \bibinfo{journal}{Science}
  \textbf{\bibinfo{volume}{269}}, \bibinfo{pages}{198} (\bibinfo{year}{1995}).

\bibitem[{\citenamefont{Davis et~al.}(1995)\citenamefont{Davis, Mewes, Andrews,
  van Druten, Durfee, Kurn, and Ketterle}}]{Davis1995}
\bibinfo{author}{\bibfnamefont{K.~B.} \bibnamefont{Davis}},
  \bibinfo{author}{\bibfnamefont{M.-O.} \bibnamefont{Mewes}},
  \bibinfo{author}{\bibfnamefont{M.~R.} \bibnamefont{Andrews}},
  \bibinfo{author}{\bibfnamefont{N.~J.} \bibnamefont{van Druten}},
  \bibinfo{author}{\bibfnamefont{D.~S.} \bibnamefont{Durfee}},
  \bibinfo{author}{\bibfnamefont{D.~M.} \bibnamefont{Kurn}}, \bibnamefont{and}
  \bibinfo{author}{\bibfnamefont{W.}~\bibnamefont{Ketterle}},
  \bibinfo{journal}{Phys. Rev. Lett.} \textbf{\bibinfo{volume}{75}},
  \bibinfo{pages}{3969} (\bibinfo{year}{1995}).

\bibitem[{\citenamefont{{Ko\u{s}trun} et~al.}(2000)\citenamefont{{Ko\u{s}trun},
  Mackie, {C\^ot\'e}, and Javanainen}}]{Kostrun2000}
\bibinfo{author}{\bibfnamefont{M.}~\bibnamefont{{Ko\u{s}trun}}},
  \bibinfo{author}{\bibfnamefont{M.}~\bibnamefont{Mackie}},
  \bibinfo{author}{\bibfnamefont{R.}~\bibnamefont{{C\^ot\'e}}},
  \bibnamefont{and}
  \bibinfo{author}{\bibfnamefont{J.}~\bibnamefont{Javanainen}},
  \bibinfo{journal}{Phys. Rev. A} \textbf{\bibinfo{volume}{62}},
  \bibinfo{pages}{063616} (\bibinfo{year}{2000}).

\bibitem[{\citenamefont{Heinzen et~al.}(2000)\citenamefont{Heinzen, Wynar,
  Drummond, and Kheruntsyan}}]{Heinzen2001}
\bibinfo{author}{\bibfnamefont{D.~J.} \bibnamefont{Heinzen}},
  \bibinfo{author}{\bibfnamefont{R.}~\bibnamefont{Wynar}},
  \bibinfo{author}{\bibfnamefont{P.~D.} \bibnamefont{Drummond}},
  \bibnamefont{and} \bibinfo{author}{\bibfnamefont{K.~V.}
  \bibnamefont{Kheruntsyan}}, \bibinfo{journal}{Phys. Rev. Lett.}
  \textbf{\bibinfo{volume}{84}}, \bibinfo{pages}{5029} (\bibinfo{year}{2000}).

\bibitem[{\citenamefont{Timmermans et~al.}(1999)\citenamefont{Timmermans,
  Tommasini, {C\^ot\'e}, Hussein, and Kerman}}]{Timmermans1999}
\bibinfo{author}{\bibfnamefont{E.}~\bibnamefont{Timmermans}},
  \bibinfo{author}{\bibfnamefont{P.}~\bibnamefont{Tommasini}},
  \bibinfo{author}{\bibfnamefont{R.}~\bibnamefont{{C\^ot\'e}}},
  \bibinfo{author}{\bibfnamefont{M.}~\bibnamefont{Hussein}}, \bibnamefont{and}
  \bibinfo{author}{\bibfnamefont{A.}~\bibnamefont{Kerman}},
  \bibinfo{journal}{Phys. Rev. Lett.} \textbf{\bibinfo{volume}{83}},
  \bibinfo{pages}{2691} (\bibinfo{year}{1999}).

\bibitem[{\citenamefont{Julienne et~al.}(1998)\citenamefont{Julienne, Burnett,
  Band, and Stwalley}}]{Julienne1998}
\bibinfo{author}{\bibfnamefont{P.~S.} \bibnamefont{Julienne}},
  \bibinfo{author}{\bibfnamefont{K.}~\bibnamefont{Burnett}},
  \bibinfo{author}{\bibfnamefont{Y.~B.} \bibnamefont{Band}}, \bibnamefont{and}
  \bibinfo{author}{\bibfnamefont{W.~C.} \bibnamefont{Stwalley}},
  \bibinfo{journal}{Phys. Rev. A} \textbf{\bibinfo{volume}{58}},
  \bibinfo{pages}{R797} (\bibinfo{year}{1998}).

\bibitem[{\citenamefont{Helmerson and You}(2001)}]{Helmerson2001}
\bibinfo{author}{\bibfnamefont{K.}~\bibnamefont{Helmerson}} \bibnamefont{and}
  \bibinfo{author}{\bibfnamefont{L.}~\bibnamefont{You}},
  \bibinfo{journal}{Phys. Rev. Lett.} \textbf{\bibinfo{volume}{87}},
  \bibinfo{pages}{170402} (\bibinfo{year}{2001}).

\bibitem[{\citenamefont{Wynar et~al.}(2000)\citenamefont{Wynar, Freeland, Han,
  Ryu, and Heinzen}}]{Wynar2000}
\bibinfo{author}{\bibfnamefont{R.}~\bibnamefont{Wynar}},
  \bibinfo{author}{\bibfnamefont{R.~S.} \bibnamefont{Freeland}},
  \bibinfo{author}{\bibfnamefont{D.~J.} \bibnamefont{Han}},
  \bibinfo{author}{\bibfnamefont{C.}~\bibnamefont{Ryu}}, \bibnamefont{and}
  \bibinfo{author}{\bibfnamefont{D.~J.} \bibnamefont{Heinzen}},
  \bibinfo{journal}{Science} \textbf{\bibinfo{volume}{287}},
  \bibinfo{pages}{1016} (\bibinfo{year}{2000}).

\bibitem[{\citenamefont{Gerton et~al.}(2000)\citenamefont{Gerton, Strekalov,
  Prodan, and Hulet}}]{Gerton2000}
\bibinfo{author}{\bibfnamefont{J.~M.} \bibnamefont{Gerton}},
  \bibinfo{author}{\bibfnamefont{D.}~\bibnamefont{Strekalov}},
  \bibinfo{author}{\bibfnamefont{I.}~\bibnamefont{Prodan}}, \bibnamefont{and}
  \bibinfo{author}{\bibfnamefont{R.~G.} \bibnamefont{Hulet}},
  \bibinfo{journal}{Nature} \textbf{\bibinfo{volume}{408}},
  \bibinfo{pages}{692} (\bibinfo{year}{2000}).

\bibitem[{\citenamefont{Weiner et~al.}(1999)\citenamefont{Weiner, Bagnato,
  Zilio, and Julienne}}]{Weiner1999}
\bibinfo{author}{\bibfnamefont{J.}~\bibnamefont{Weiner}},
  \bibinfo{author}{\bibfnamefont{V.~S.} \bibnamefont{Bagnato}},
  \bibinfo{author}{\bibfnamefont{S.}~\bibnamefont{Zilio}}, \bibnamefont{and}
  \bibinfo{author}{\bibfnamefont{P.~S.} \bibnamefont{Julienne}},
  \bibinfo{journal}{Rev. Mod. Phys.} \textbf{\bibinfo{volume}{71}},
  \bibinfo{pages}{1} (\bibinfo{year}{1999}).

\bibitem[{\citenamefont{Kozuma et~al.}(1999)\citenamefont{Kozuma, Deng, Hagley,
  Wen, Lutwak, Helmerson, Rolston, and Phillips}}]{Kozuma1999}
\bibinfo{author}{\bibfnamefont{M.}~\bibnamefont{Kozuma}},
  \bibinfo{author}{\bibfnamefont{L.}~\bibnamefont{Deng}},
  \bibinfo{author}{\bibfnamefont{E.~W.} \bibnamefont{Hagley}},
  \bibinfo{author}{\bibfnamefont{J.}~\bibnamefont{Wen}},
  \bibinfo{author}{\bibfnamefont{R.}~\bibnamefont{Lutwak}},
  \bibinfo{author}{\bibfnamefont{K.}~\bibnamefont{Helmerson}},
  \bibinfo{author}{\bibfnamefont{S.~L.} \bibnamefont{Rolston}},
  \bibnamefont{and} \bibinfo{author}{\bibfnamefont{W.~D.}
  \bibnamefont{Phillips}}, \bibinfo{journal}{Phys. Rev. Lett.}
  \textbf{\bibinfo{volume}{82}}, \bibinfo{pages}{871} (\bibinfo{year}{1999}).

\bibitem[{\citenamefont{Andrews et~al.}(1996)\citenamefont{Andrews, Mewes, van
  Druten, Durfee, Kurn, and Ketterle}}]{Andrews1996}
\bibinfo{author}{\bibfnamefont{M.~R.} \bibnamefont{Andrews}},
  \bibinfo{author}{\bibfnamefont{M.-O.} \bibnamefont{Mewes}},
  \bibinfo{author}{\bibfnamefont{N.~J.} \bibnamefont{van Druten}},
  \bibinfo{author}{\bibfnamefont{D.~S.} \bibnamefont{Durfee}},
  \bibinfo{author}{\bibfnamefont{D.~M.} \bibnamefont{Kurn}}, \bibnamefont{and}
  \bibinfo{author}{\bibfnamefont{W.}~\bibnamefont{Ketterle}},
  \bibinfo{journal}{Science} \textbf{\bibinfo{volume}{273}},
  \bibinfo{pages}{84} (\bibinfo{year}{1996}).

\bibitem[{\citenamefont{Jones et~al.}(2000)\citenamefont{Jones, Lett, Tiesinga,
  and Julienne}}]{Jones2000}
\bibinfo{author}{\bibfnamefont{K.~M.} \bibnamefont{Jones}},
  \bibinfo{author}{\bibfnamefont{P.~D.} \bibnamefont{Lett}},
  \bibinfo{author}{\bibfnamefont{E.}~\bibnamefont{Tiesinga}}, \bibnamefont{and}
  \bibinfo{author}{\bibfnamefont{P.~S.} \bibnamefont{Julienne}},
  \bibinfo{journal}{Phys. Rev. A} \textbf{\bibinfo{volume}{61}},
  \bibinfo{pages}{012501} (\bibinfo{year}{2000}).

\bibitem[{\citenamefont{Samuelis et~al.}(2001)\citenamefont{Samuelis, Tiesinga,
  Laue, Elbs, Kn{\"o}ckel, and Tiemann}}]{Samuelis2001}
\bibinfo{author}{\bibfnamefont{C.}~\bibnamefont{Samuelis}},
  \bibinfo{author}{\bibfnamefont{E.}~\bibnamefont{Tiesinga}},
  \bibinfo{author}{\bibfnamefont{T.}~\bibnamefont{Laue}},
  \bibinfo{author}{\bibfnamefont{M.}~\bibnamefont{Elbs}},
  \bibinfo{author}{\bibfnamefont{H.}~\bibnamefont{Kn{\"o}ckel}},
  \bibnamefont{and} \bibinfo{author}{\bibfnamefont{E.}~\bibnamefont{Tiemann}},
  \bibinfo{journal}{Phys. Rev. A} \textbf{\bibinfo{volume}{63}},
  \bibinfo{pages}{012710} (\bibinfo{year}{2001}).

\bibitem[{\citenamefont{Jones et~al.}(1997)\citenamefont{Jones, Maleki,
  Ratliff, and Lett}}]{Jones1997}
\bibinfo{author}{\bibfnamefont{K.~M.} \bibnamefont{Jones}},
  \bibinfo{author}{\bibfnamefont{S.}~\bibnamefont{Maleki}},
  \bibinfo{author}{\bibfnamefont{L.~P.} \bibnamefont{Ratliff}},
  \bibnamefont{and} \bibinfo{author}{\bibfnamefont{P.~D.} \bibnamefont{Lett}},
  \bibinfo{journal}{J. Phys. B} \textbf{\bibinfo{volume}{30}},
  \bibinfo{pages}{289} (\bibinfo{year}{1997}).

\bibitem[{\citenamefont{Gerton et~al.}(2001)\citenamefont{Gerton, Frew, and
  Hulet}}]{Gerton2001}
\bibinfo{author}{\bibfnamefont{J.~M.} \bibnamefont{Gerton}},
  \bibinfo{author}{\bibfnamefont{B.~J.} \bibnamefont{Frew}}, \bibnamefont{and}
  \bibinfo{author}{\bibfnamefont{R.~G.} \bibnamefont{Hulet}},
  \bibinfo{journal}{Phys. Rev. A} \textbf{\bibinfo{volume}{64}},
  \bibinfo{pages}{053410} (\bibinfo{year}{2001}).

\bibitem[{\citenamefont{Bohn and Julienne}(1999)}]{Bohn1999}
\bibinfo{author}{\bibfnamefont{J.~L.} \bibnamefont{Bohn}} \bibnamefont{and}
  \bibinfo{author}{\bibfnamefont{P.~S.} \bibnamefont{Julienne}},
  \bibinfo{journal}{Phys. Rev. A} \textbf{\bibinfo{volume}{60}},
  \bibinfo{pages}{414} (\bibinfo{year}{1999}).

\bibitem[{\citenamefont{Gallagher and Pritchard}(1989)}]{Gallagher1989}
\bibinfo{author}{\bibfnamefont{A.}~\bibnamefont{Gallagher}} \bibnamefont{and}
  \bibinfo{author}{\bibfnamefont{D.}~\bibnamefont{Pritchard}},
  \bibinfo{journal}{Phys. Rev. Lett.} \textbf{\bibinfo{volume}{63}},
  \bibinfo{pages}{957} (\bibinfo{year}{1989}).

\bibitem[{\citenamefont{Bohn and Julienne}(1996)}]{Bohn1996}
\bibinfo{author}{\bibfnamefont{J.~L.} \bibnamefont{Bohn}} \bibnamefont{and}
  \bibinfo{author}{\bibfnamefont{P.~S.} \bibnamefont{Julienne}},
  \bibinfo{journal}{Phys. Rev. A} \textbf{\bibinfo{volume}{54}},
  \bibinfo{pages}{R3647} (\bibinfo{year}{1996}).

\bibitem[{\citenamefont{Javanainen and Mackie}(2001)}]{Javanainen2001}
\bibinfo{author}{\bibfnamefont{J.}~\bibnamefont{Javanainen}} \bibnamefont{and}
  \bibinfo{author}{\bibfnamefont{M.}~\bibnamefont{Mackie}},
  \bibinfo{howpublished}{cond-mat/0108349} (\bibinfo{year}{2001}).

\bibitem[{\citenamefont{Javanainen}()}]{Javanainen2001b}
\bibinfo{author}{\bibfnamefont{J.}~\bibnamefont{Javanainen}},
  \bibinfo{howpublished}{private communication}.

\end{thebibliography}

\end{document}